# Modeling seismic wave propagation and amplification in 1D/2D/3D linear and nonlinear unbounded media

J. F. Semblat
*Dept. of Soil and Rock Mechanics, LCPC, University Paris-East, Paris, France*

**ABSTRACT:** To analyze seismic wave propagation in geological structures, it is possible to consider various numerical approaches: the finite difference method, the spectral element method, the boundary element method, the finite element method, the finite volume method, etc. All these methods have various advantages and drawbacks. The amplification of seismic waves in surface soil layers is mainly due to the velocity contrast between these layers and, possibly, to topographic effects around crests and hills. The influence of the geometry of alluvial basins on the amplification process is also know to be large. Nevertheless, strong heterogeneities and complex geometries are not easy to take into account with all numerical methods. 2D/3D models are needed in many situations and the efficiency/accuracy of the numerical methods in such cases is in question. Furthermore, the radiation conditions at infinity are not easy to handle with finite differences or finite/spectral elements whereas it is explicitely accounted in the Boundary Element Method. Various absorbing layer methods (e.g. F-PML, M-PML) were recently proposed to attenuate the spurious wave reflections especially in some difficult cases such as shallow numerical models or grazing incidences. Finally, strong earthquakes involve nonlinear effects in surficial soil layers. To model strong ground motion, it is thus necessary to consider the nonlinear dynamic behaviour of soils and simultaneously investigate seismic wave propagation in complex 2D/3D geological structures! Recent advances in numerical formulations and constitutive models in such complex situations are presented and discussed in this paper. A crucial issue is the availability of the field/laboratory data to feed and validate such models.

## 1 Modeling seismic wave propagation

Many various numerical methods are available to model seismic wave propagation and we will discuss them first. Afterwards, the issue of seismic wave amplification (site effects) in both linear (weak motion) and nonlinear (strong motion) ranges will then be examined.

To analyze seismic wave propagation in 2D or 3D geological structures, various numerical methods are available (Fig.1):
- the finite difference method is accurate in elastodynamics but is mainly adapted to simple geometries (Bohlen, 2006, Frankel 1992, Moczo 2002, Virieux 1986),
- the finite element method is efficient to deal with complex geometries and numerous heterogeneities (even for inelastic constitutive models (Bonilla, 2000)) but has several drawbacks such as numerical dispersion and numerical damping (Hughes 1987, 2008, Ihlenburg 1995, Semblat, 2000a, 2008,) and (consequently) numerical cost in 3D elastodynamics,
- the spectral element method has been increasingly considered to analyse 2D/3D wave propagation in linear media with a good accuracy due to its spectral convergence properties (Chaljub, 2007, Faccioli, 1996, Komatitsch, 1998),
- the boundary element method allows a very good description of the radiation conditions but is preferably dedicated to weak heterogeneities and linear constitutive models (Beskos 1997, Bonnet 1999, Dangla 1988, 2005, Sanchez-Sesma 1995, Semblat 2008, 2000b). Recent developments have been proposed to reduce the computational cost of the method especially in the high frequency range (Chaillat, 2008, 2009, Fujiwara, 2000),
- the finite volume method was recently developed in the field of elastodynamics (Glinsky, 2006),
- the Aki-Larner method which takes advantage of the frequency-wavenumber decomposition but is limited to simple geometries (Aki 1970, Bouchon 1989),
- the scaled boundary finite element method which is a kind of solution-less boundary element method (Wolf, 2003),
- other methods such as series expansions of wave functions (Liao 2004, Sanchez-Sesma 1983).

Furthermore, when dealing with wave propagation in unbounded domains, many of these numerical methods raise the need for absorbing boundary conditions to avoid spurious reflections. Since each method has specific advantages and shortcomings (Table I), it is consequently often more interesting to combine two methods to take advantage of their peculiarities. It is for instance possible to couple FEM and BEM (Aochi 2005, Dangla 1988, Bonnet 1999) allowing an accurate description of the near field (FEM model including complex geometries, numerous heterogeneities and nonlinear constitutive laws) and a reliable estimation of the far-field (BEM involving accurate radiation conditions).

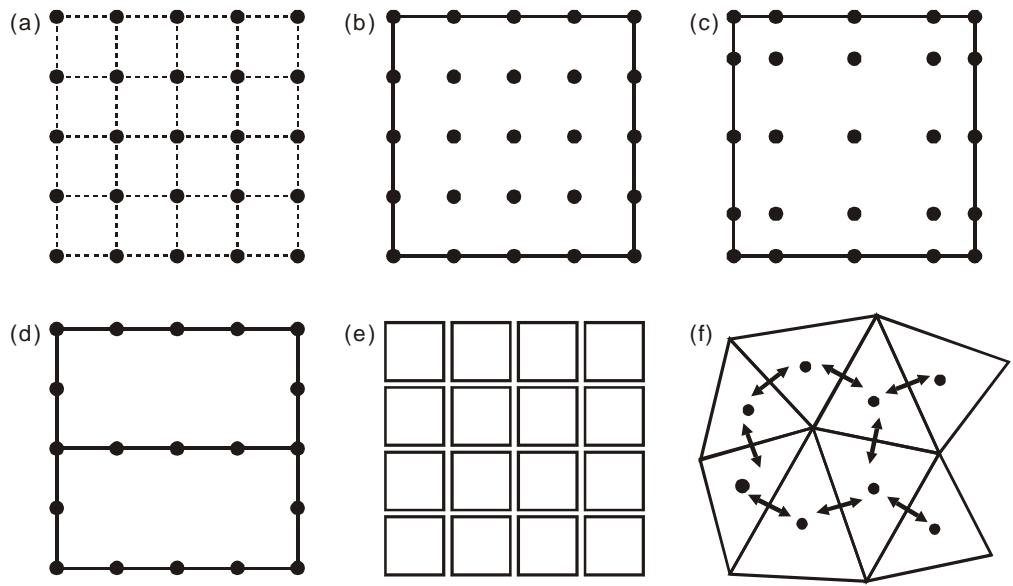

Figure 1. Various numerical methods to model seismic wave propagation: (a) the Finite Difference Method, (b) the Finite Element Method, (c), the Spectral Element Method, (d) the Boundary Element Method, (e) the Discrete Element Method, (f) the Finite Volume Method.

Table I. Features of various numerical methods for wave propagation modeling.

| Method | FDM | FEM | SEM | BEM | DEM | FVM |
|---|---|---|---|---|---|---|
| Continuous | yes | yes | yes | yes | no(/yes) | yes |
| Constitutive | lin./NL | lin./NL | lin./NL | linear | lin./NL | lin. |
| Radiation | abs. bound. | abs. bound. | abs. bound. | intrinsic | abs. bound. | abs. bound. |

## 2 Modeling seismic wave propagation by the Boundary Element Method

### 2.1 Interest of the Boundary Element Method ("BEM")

The main advantage of the method in elastodynamics is the accurate description of the infinite extension of the medium through exact radiation conditions (Bonnet 1999, Semblat 2008). In contrast to other discretization methods, there is no need to consider absorbing boundary conditions.

The Boundary Element Method involves singular integrals (Bonnet 1999). There are three main kinds of singularities: (i) weak singularity (i.e. in the ordinary Riemann sense), (ii) strong singularity (i.e. in the Cauchy principal value sense) or (iii) hyper-singularity (i.e. in the Hadamard finite part sense). Strong and hyper-singular integrals have to be converted to regular ones in the regularization of the BEM formulations (Bonnet 1999, Sladek 1998). Indeed, weak singularity is not treated by regularization. However, from the point of view of numerical integrations, one should devote a great attention to the evaluation of these integrals because standard integration quadratures fail in accuracy (Beskos 1997, Dangla 1988). Therefore each type of singularity has to be treated by appropriate techniques (Bonnet 1993, Dangla 2005, Sladek 1998). It is noticed that the regularization can be performed either before or after the discretization, i.e. in the global or local (intrinsic) coordinate space, as observed in some papers mentioned above. A comprehensive review of BEM in dynamic analysis has been proposed by Beskos (1997). An example (pressurized cavity in a full space) showing the influence of the regularization process on the numerical solution is proposed in Fig. 2: the uncorrected solution is very far from the regularized one which matches well the analytical solution (Dangla, 2005).

The main drawback of the BEM is that it leads to full non symmetric matrix systems. For large models, the numerical cost and memory storage requirements may then by huge. Recent researches considered symmetric Galerkin boundary element methods more efficient for large numbers of unknowns (Bonnet 1998). Current researches try to reduced the number of computations through Fast Multipole formulations initially developed in the field of physics (Fujiwara, 2000, Chaillat 2008, 2009). Such alternative approaches are discussed in the following.

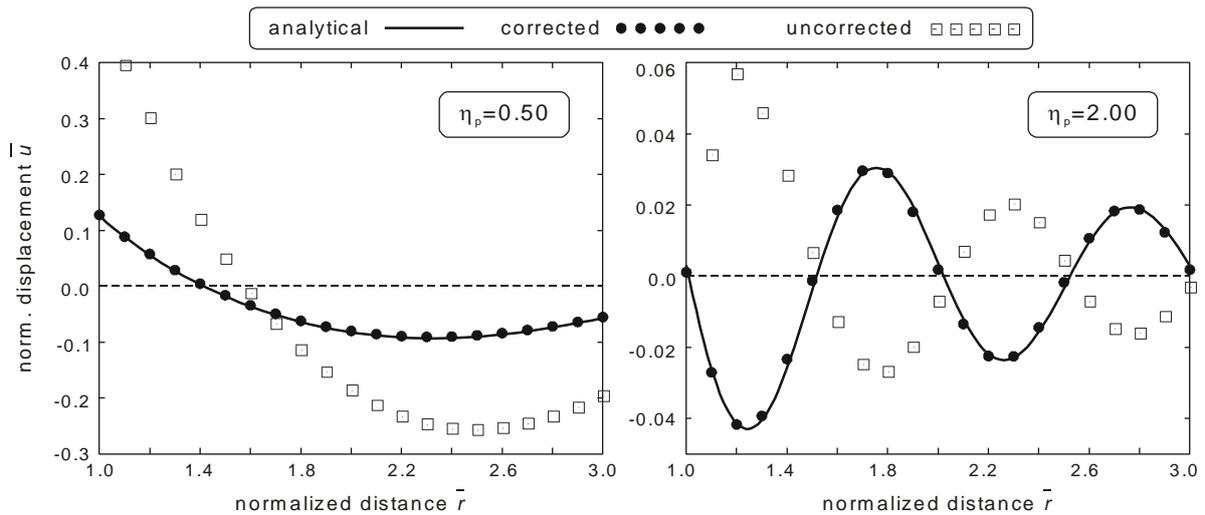

Figure 2. Comparison of regularized and uncorrected BEM solutions for a pressurized cavity in a full-space: harmonic loading on the cavity wall at two different normalized frequencies $\eta=2R/\lambda_P$ (Dangla, 2005).

### 2.2 Fast Multipole formulation for the Boundary Element Method

In the field of physics (Maxwell or Laplace equations), recent advances in boundary element methods lead to a very important decrease of the computational cost. Instead of point to point interaction as in classical boundary element methods, the fast multipole method (Greengard 1998, Fujiwara 2000) considers interactions between groups of points (cells centered on a multipole, Fig. 3) and hence avoids multiple computations of nearly identical terms corresponding to very close points.

It is possible to apply this method at a single scale or even at various scales through a multilevel approach. The size of each cell around its multipole depends on the distance to the other cells: the larger the distance between cells, the larger the cells (Fig.3). The computations of the singular integrals are then performed through this approach. They are split in an integral on the surface around the singularity and another integral on the complementary distant surface. The first one is estimated using classical regularization techniques, whereas the latter is computed with a fast multipole algorithm (Greengard 1998). The advantage for the computational cost is very significant since it depends on $N^2$ for classical methods and on $N$ or $N \log N$ for the fast multipole method (Chaillat 2008). Furthermore, the computational cost is reduced for both the memory storage and the calculation time. This method allows the analysis of very large problems involving millions of unknowns on a single-processor PC (to compare to several tens of thousands previously).

Current researches also investigate a fast multipole method well-adapted to the Helmholtz equation and even to elastodynamics (Fujiwara 2000, Chaillat 2008, 2009). The fast multipole method then allows the computation of very large models considering a larger number of heterogeneities, a more realistic geometrical representation of geological structures (especially in 3D) as well as higher frequency values (detailed modeling of short wavelengthes amplification).

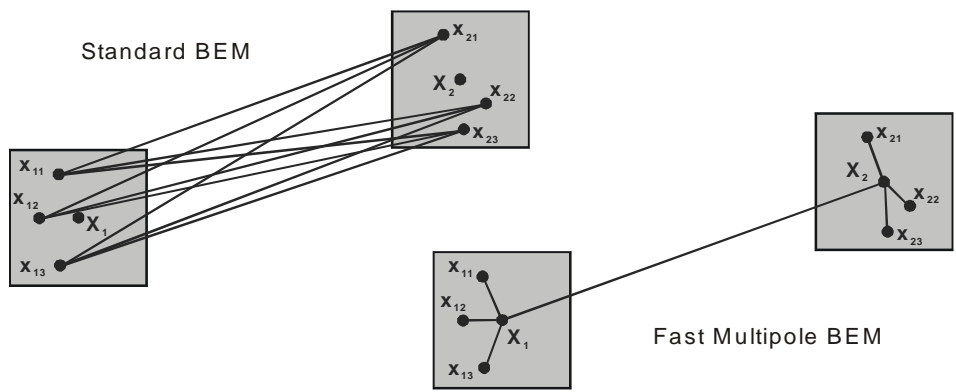

Figure 3. Comparison of the principles of the standard BEM and the Fast Multipole BEM.

## 2.3 Modeling seismic wave amplification by the Boundary Element Method

*2.3.1 Amplification of seismic waves in geological structures*

The amplification of the seismic motion mainly occurs in alluvial deposits the characteristics of which (geometry, wave velocities) control the amplification process (Bard 1985, Dobry 1976, Duval 1998, Sanchez-Sesma 1983, 2000, Semblat 2000, 2003). Considering the Boundary Element Method, seismic wave amplification (i.e. site effects) can be analyzed numerically taking into account the infinite extension of the medium as well as complex geometrical features. The main phenomena that can be recovered by the BEM models are the following:

- amplification due to the (1D) velocity contrast between soil layers,
- focusing effects due to complex layers or free-surface geometries,
- basins effects (2D/3D) due the trapped surfaces waves in surficial layers.

Significant wave scattering may also be found around strong topographic irregularities such as crests or hills slopes (Bouckovalas 2005, Paolucci 2002, Reinoso 1997, Sanchez-Sesma 1983, Semblat 2009). Such phenomena are called topographic site effects and may aso be studied by the Boundary Element Method.

In both case (alluvial basins and topographic irregularities), BEM models in 2D or 3D allow the estimation of the amplification factor of the seismic motion. The BEM formulation can be considered in frequency-domain for long durations and narrow bandwidths (Bonnet 1999, Dangla 2005) or time-domain for shorter durations and larger bandwidths (Gaul 1999, Jin 2001, Manolis 1988). Other numerical methods may also be considered as discussed in previous sections (Faccioli 1996, Glinsky 2006, Komatitsch 1998, Moczo 2002, Virieux 1986).

*2.3.2 Topographic amplification of seismic waves*

The analysis of wave scattering around topographic irregularities has been performed by various authors for canonical configurations (Bouckovalas 2005, Reinoso 1997, Sanchez-Sesma 1983) or for actual 2D/3D topographies (Paolucci 2002, Semblat 2002). As shown in Fig. 4 in the case of Caracas (Semblat 2002), the maximum amplification $A_{max}$ along a complex topography may lead to significant changes in the seismic wavefield. When comparing the top plot at frequency 0.25 Hz to the bottom one (0.35 Hz), these topographic site effects are nevertheless found to be less than those found in alluvial basins. Detailed analyses of such effects are proposed in (Bouckovalas 2005, Paolucci 2002, Semblat 2009). Bouckovalas and Papadimitriou (2005) proposed empirical laws to estimate the topographic effects for 2D steep crests.

For 3D models, efficient numerical methods are needed. Paolucci (2002) studied an actual steep italian site using the Spectral Element Method. Since the development of Fast Multipole BEMs in elastodynamics allows to handle large scale 3D models, several large canonical 3D topographies have been recently studied (Chaillat 2008). As shown in Figure 5 for an ellipsoidal canyon and an oblique incident plane P-wave (Chaillat 2008), the FM-BEM solution is in good agreement with previous solutions obtained by Reinoso (1997) for both the vertical and horizontal motion components. Several other canonical configurations were defined in the framework of the *QSHA* research project (*Quantitative Seismic Hazard Assessment*) and are used to compare the efficiency and accuracy of various numerical methods (data available at http://qsha.unice.fr/). Current researches aim at investigating actual 3D topographies by the Fast Multipole Method.

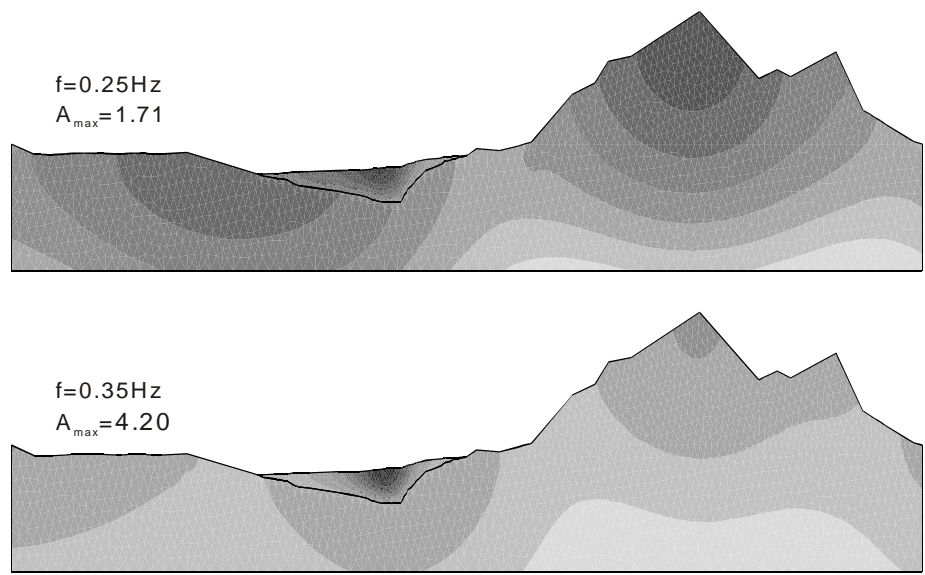

Figure 4. Topographic (top) vs statigraphic (bottom) seismic wave amplification: example in Caracas, Venezuela (Semblat et al., 2002).

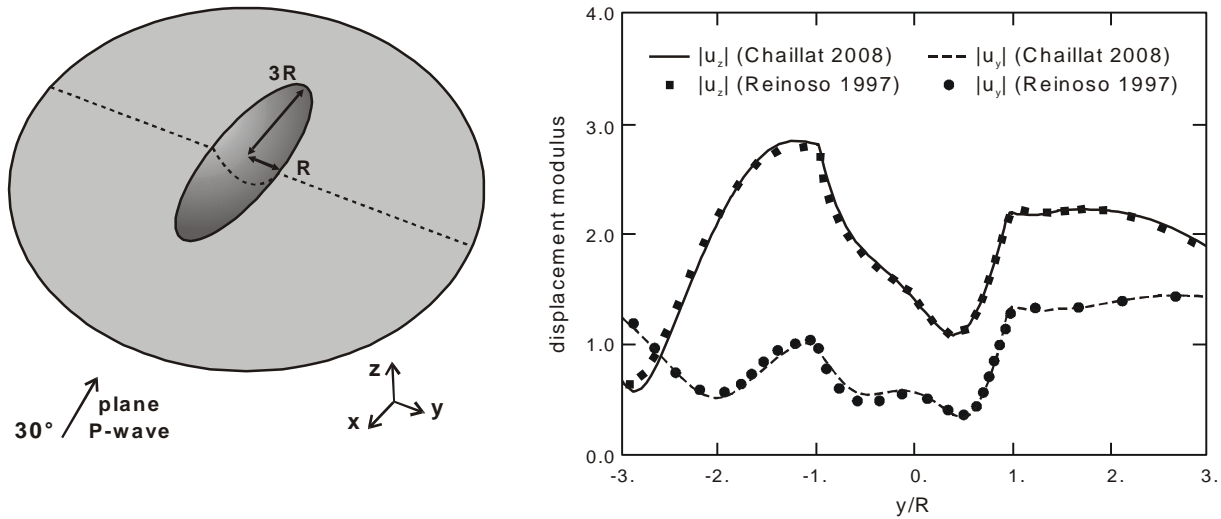

Figure 5. Validation of the Fast Multipole BEM from a 3D configuration studied by Reinoso (1997): diffraction of an oblique plane P-wave by a semi-ellipsoidal canyon (Chaillat 2008).
(data for various canonical configurations available at http://qsha.unice.fr/)

*2.3.3 Amplification of seismic waves in alluvial basins*

The amplification of seismic waves in alluvial basins is generally larger than that observed/computed around topographic irregularities (Baise 2003, Beauval 2003, Frankel 1992, Sanchez-Sesma 1995, Semblat 2009, Sommerville 1998). Since extensive field measurements were performed on Volvi EuroSeisTest, Greece (Bard 2000, Beauval 2003, Chavez-Garcia 2000, Pitilakis 1999), BEM models were performed for this site (Semblat 2005). The numerical results were compared to numerous seismological field data from various earthquake recordings.

The Volvi basin is 6km wide and 250m deep. As shown in Figure 6, the amplification level changes from one frequency to the other. The influence on the amplification process of both the basin geometry and the soil layering were studied in (Semblat 2005). Since the geometry of the basin is complex, strong focusing effects occur leading to large amplification levels (around 10) at some peculiar frequencies (Figure 6). However, depending on the accuracy of the geological model, the description of the soil layering has a strong influence at higher frequencies (Figure 6, bottom right). It has been quantified for a simplified (2 layers) as well as detailed (6 layers) model of the Volvi profile in (Semblat 2005). The detailed analysis leads to larger amplitude trapped surface waves at higher frequencies.

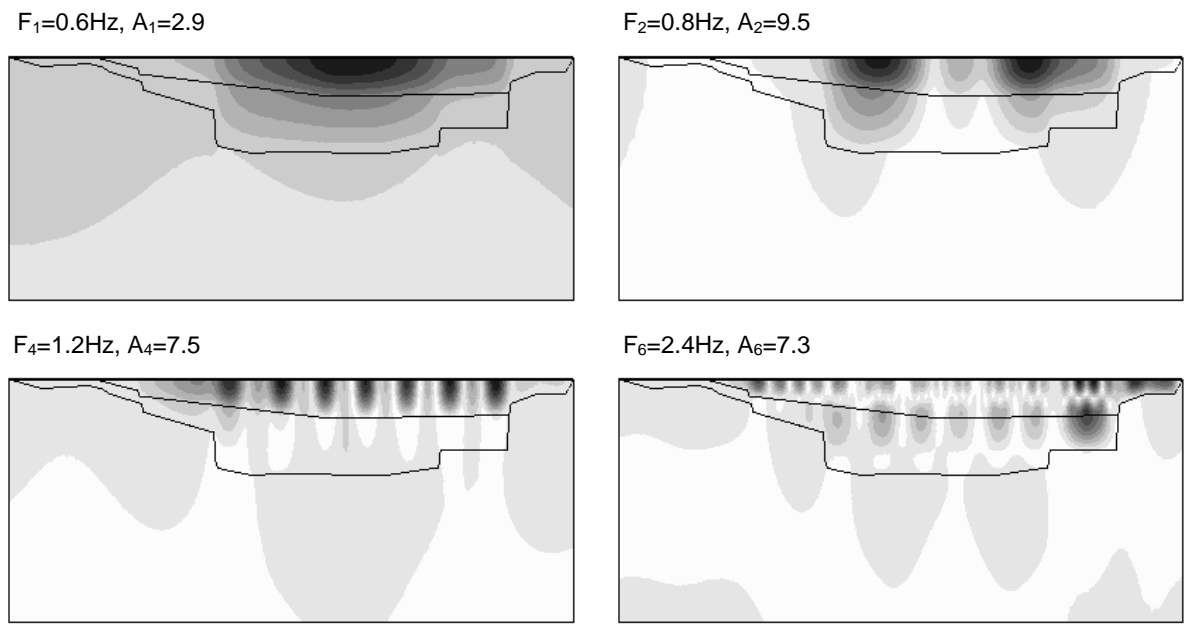

Figure 6. Seismic wave amplification in the Volvi-EuroSeisTest basin, Greece (Semblat 2005): BEM simulations at different frequencies $F_i$ and maximum related amplifications.

## 3 Modeling seismic wave propagation by the Finite/Spectral Element Method

### 3.1 Numerical wave dispersion

Considering the finite/spectral element method to model seismic wave propagation, as for other numerical methods, two types of numerical errors can be studied (Hughes, 1987, Chaljub, 2007, Ihlenburg, 1995) : the relative period error and the algorithmic damping. The relative period error is different from one integration scheme to another (Hughes, 1987). For wave propagation problems, the relative period error appears in the estimation of wave-velocities and is called numerical dispersion. Seismic wave propagation in a numerical scheme then depends on the element size, the integration scheme, the element type, etc. This phenomenon is called numerical dispersion as a reference to physical dispersion leading to a dependence of the wave-velocity on frequency (Deraemaeker 1999, Hughes 2008, Ihlenburg, 1995, Semblat 2000b). To model wave propagation phenomena by the finite element method or the finite difference method, one has to control the accuracy of the numerical scheme since the numerical error tends to increase during the propagation process.

Various theoretical works concern the analysis of the numerical error made in the estimation of the approximated wave number (Deraemaeker 1999, Hughes 2008, Semblat 2009). For instance, it was shown that there is a cut-off frequency above which there is no propagation at all (Ihlenburg, 1995). Depending on the frequency of the excitation, the numerical wave propagates slower or faster than the theoretical solution. It is then necessary to analyze numerical dispersion of seismic waves and precisely quantify the numerical error.

### 3.2 Higher order finite elements vs spectral elements

Numerical wave dispersion is also influenced by the order of the shape functions of the finite elements. Higher order finite elements are for instance known to have a very good precision for elastic-plastic computations. In acoustics and elastodynamics, several theoretical works propose analytical expressions to estimate the numerical dispersion (Deraemaeker 1999, Ihlenburg, 1995). Some comparisons between lower order and higher order finite elements to model wave propagation were recently proposed by Hughes (2008).

In Figure 7, the results from 1D FEM simulations (Semblat 2000b) are plotted for various shape functions orders. The same number of degrees of freedom is considered in each case. The linear finite elements (Figure 7, top) are shown to have strong numerical dispersion (numerical wave velocity artificially increased). For quadratic finite elements (Figure 7, middle), the accuracy is satisfactory when compared to the theoretical delays (vertical dashed lines). Finally, for the higher order finite elements (Figure 7, bottom), the numerical wave dispersion is very low. These results show that, for the same number of DOFs, the accuracy of higher order finite element for wave propagation simulations is much better.

In addition to "classical" higher order finite elements, spectral finite elements have increasingly been studied since their spectral convergence is of great interest for wave propagation simulations (Chaljub 2007, Faccioli 1996, Komatitsch 1998). The spectral elements are generally considered for high order only (4 to 8) since the difference with classical finite elements is not significant below. An evaluation of the cost effectiveness (and accuracy) of both Chebyshev spectral and p-finite elements has been proposed by Dauksher (1999).

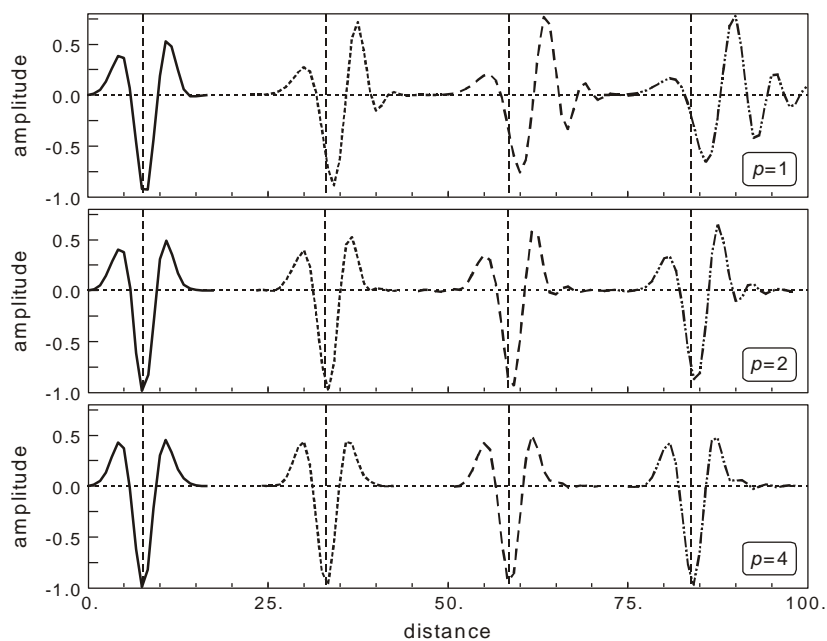

Figure 7. Numerical wave dispersion for various finite element orders: linear (top), quadratic (middle) and fourth-order (bottom) (Semblat 2000b).

### 3.3 Modeling linear attenuation

The analysis of seismic wave propagation in attenuating media, such as soils and rocks, raises the need for suitable attenuation models. One of the most common damping model in structural dynamics is the Rayleigh formulation giving a damping matrix as a linear combination of stiffness and mass matrices (Hughes 1987). The damping ratio is then minimum for a given frequency and infinite for zero and infinite frequencies (Semblat 2009). In the field of structural dynamics, the main advantage of the Rayleigh formulation is that it leads to a diagonal damping marix in the real eigenmodes base (Chopra 2007).

For wave propagation phenomena, the Rayleigh damping formulation is equivalent to a particular rheological model: generalized Maxwell model (Semblat 1997). For weak to moderate damping values, this equivalence allows an easier and explicit estimation of the two coefficients of the Rayleigh formulation from experimental behaviour parameters. Other types of "matrix based" damping formulations are available such as the Caughey formulation (Chopra 2007, Semblat 2009) corresponding to a generalization of the Rayleigh formulation and allowing complex damping-frequency dependences.

Various types of mechanical or rheological viscoelastic formulations were also proposed. Kjartansson (1979) investigated frequency constant attenuation ("CQ") models for wave propagation problems. Day and Minster (1984) considered a Padé approximant to simulate wave propagation in attenuatingmedia. Emmerich and Korn (1987) proposed a rheological model leading to a nearly constant attenuation ("NCQ") in a given frequency range. Carcione et al. (2002) studied a damping formulation based on fractional derivatives and Moczo and Kristek (2005) discussed time domain simulations using rheological models. All these approaches are formulated in the linear range and need, for time domain computations, to deal with memory variables which may be costly. As discussed in the following, nonlinear constitutive laws are often necessary in the case of strong earthquakes.

## 4 Radiation conditions at infinity

For finite differences, finite or spectral element methods, it is necessary to take into account the radiation conditions through efficient techniques to avoid spurious wave reflections at the mesh boundaries. It can be done considering absorbing boundary conditions or infinite elements for purely solid (Chadwick 1999) as well as multiphase media (Modaressi 1994). It is nevertheless difficult for very heterogeneous media (Chammas 2003).

Absorbing layer methods rather try to model the wave attenuation in a thick layer at the medium boundaries (Semblat 2009). They have been increasingly developed in recent years since their efficiency appears good in various configurations. They are generally known as "Perfectly Matched Layer" methods.

Various types of PML formulations have recently been proposed:

- *Classical PML* (Basu 2003): the wavefield propagating in the layer is damped according to a one-dimensional amplitude decrease law similar to that of an attenuating medium:

$$\tilde{x} = x + \frac{\Sigma(x)}{i\omega} \tag{1}$$

This complex valued law leads to an amplitude decrease of the numerical wave along the *x*-axis. As illustrated in Figure 8, grazing incidences are related to small horizontal wavenumbers and a limited amplitude decrease will be obtained. This uniaxial law is thus not sufficient to have an optimal absorbing effect in all configurations.

- *Filtering PML* (Festa 2003, 2005): for shallow numerical models, the classical PML formulation may also amplify surface waves (Figure 8). A filtering PML formulation (F-PML) has been proposed by Festa and Vilotte (2005) to avoid this problem:

$$\tilde{x} = x + \frac{\Sigma(x)}{i\omega + \omega_c} \tag{2}$$

- *Multidirectional PML* (Meza Fajardo, 2008): another alternative to ensure the numerical stability of PMLs (e.g. grazing incidences) is the multidirectional PML formulation (M-PML) recently proposed by Meza Fajardo and Papageorgiou (Meza Fajardo, 2008). The original uniaxial law is generalized in the multi-axial case through the following system:

$$\begin{cases} \tilde{x} = x + \dfrac{1}{i\omega}\alpha_x^{(x)} x \\ \tilde{y} = y + \dfrac{1}{i\omega}\alpha_y^{(x)} y \\ \tilde{z} = z + \dfrac{1}{i\omega}\alpha_z^{(x)} z \end{cases} \tag{3}$$

It is thus possible to choose the attenuation vector α in order to optimize the absorbing effect depending on the wave type and incidence. Another interest of this formulation is its numerical stability for strongly anisotropic media (Figure 8).

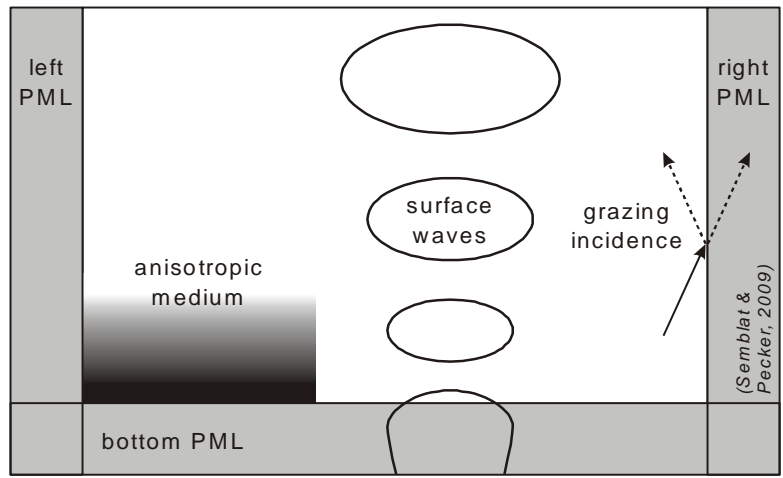

Figure 8. Various critical configurations for the efficiency/stability
of PML absorbing conditions (from Semblat, 2009).

## 5  Modeling strong ground motion and non linear effects

To model strong ground motion, it is necessary to combine seismological tools to geotechnical methods involving nonlinear constitutive laws (Heuze 2004). The nonlinear soil models should reproduce one of the main features of soil dynamic/cyclic behaviour: the shear modulus decreases and the hysteretic damping increases with strain amplitude (Seed 1986). Such models are generally considered in the framework of the finite element method and often involve Masing type cyclic behaviour (Bonilla 2000). The influence of plasticity, confining pressure and pore-pressure build up can also be analyzed with more complex models (Aubry 1982, Gyebi 1992, Lade 1977, Loret 1997, Mellal 1998, Park 2004, Prevost, 1985). It is then possible to have a deeper insight into the liquefaction processes. More and more strong motion data from well-controlled bore-holes are available and allow comparisons with nonlinear computations.

In the last decades, linear (viscoelastic) equivalent models were also extensively used to have a simplified description (i.e. few parameters) of the shear modulus decrease and the hysteretic damping increase (Schnabel 1972). Simplified models are interesting since they may allow combined seismological/geotechnical computations to simultaneously take into account basin effects at large scales (2D/3D) and nonlinear local effects. Recent researches proposed new simplified models avoiding some drawbacks of equivalent models such as frequency independence (Assimaki 2000, Kausel 2002). As shown in Fig.9 (left), this nonlinear frequency dependent model combines a strain spectrum with the classical $G(\gamma)$ and $\beta(\gamma)$ curves. Delépine (2007, 2009) proposed another alternative: a simple constitutive model generalizing the classical NCQ (Nearly Constant) attenuation model (Emmerich 1987) in the nonlinear range. This extended NCQ model leads to decreasing shear modulus and increasing damping/attenuation. This model allows the analysis of seismic waves propagation from strong earthquakes and may be used at large scales (e.g. alluvial basins). Finally, a crucial issue is the availability of the field/laboratory data to feed and validate such models.

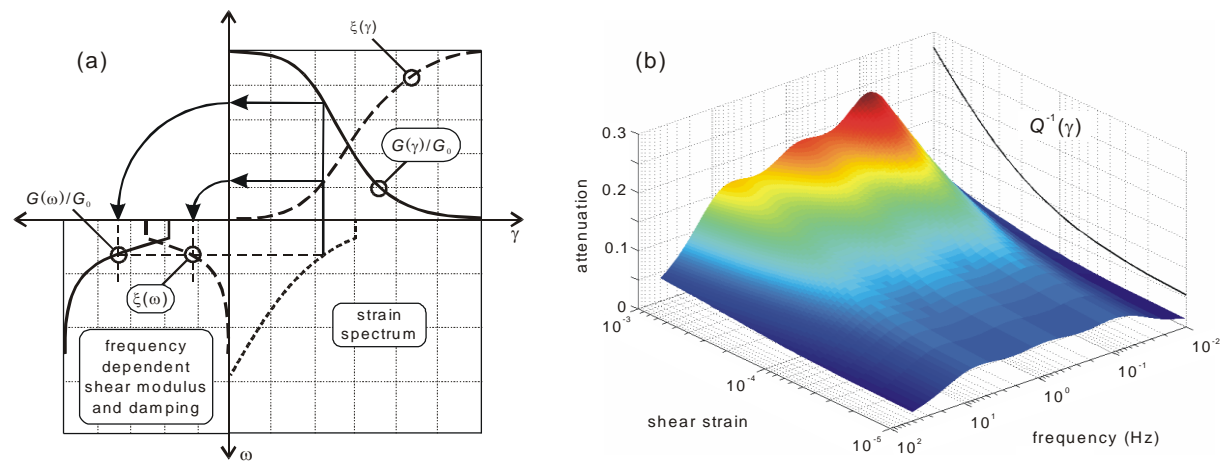

Figure 9. Recent improvements of the equivalent linear model for the analysis of strong seismic motion in soils: (a) Frequency dependent nonlinear model (Kausel 2002); (b) Nonlinear "extended NCQ" model (Delépine 2009).

# 6 Conclusion

This paper presents recent results in various research fields related to seismic wave propagation in geological structures. New methods or major improvements of existing methods (staggered grid FD, Finite Volumes, Fast Multipole-BEM) are promising ways to better model seismic wave propagation, amplification and interactions in 3D surficial geological structures. The major advances of current works also concern improved treatments of the radiation conditions for spectral elements/finite elements (absorbing layer methods such as F-PML or M-PML). Another key point is strong seismic motion: recent simplified nonlinear constitutive models allow the analysis of seismic wave propagation from strong earthquakes at large scales. Nevertheless, several topics have not been discussed in this paper: fault mechanics and the influence of fluids (Aochi 2005), soil-structure interaction (Kham 2006, Sextos 2003) and site-city interaction (Bielak 1999, Clouteau 2001, Groby 2005, Semblat 2008, Tsogka 2003, Wirgin 1996), soil-pile interaction and liquefaction (Brennan 2002, Takahashi 2005), etc.

# 7 Acknowledgements

The author would like to thank all his colleagues and students who contributed to some of the results presented herein and the related papers listed in the References.

# 8 References

Aki, K. and Larner, K.L. 1970. Surface Motion of a Layered Medium Having an Irregular Interface due to Incident Plane SH Waves, *J. Geophys. Res.*, 75, 1921-1941.

Aochi, H., Seyedi, M., Douglas, J., Foerster, E., and Modaressi, H. 2005. A complete BIEM-FDM-FEM simulation of an earthquake scenario, *General Assembly of the European Geosciences Union*.

Assimaki, D., Kausel E., Whittle, A. 2000. Model for dynamic shear modulus and damping in granular soils, *Jal of Geotechnical and Geoenvironmental Eng.*, 126(10), 859-869.

Aubry D., Hujeux J.C., Lassoudire F., Meimon Y. 1982. A double memory model with multiple mechanisms for cyclic soil behaviour, *Proc. Int. Symp. Num. Mod. Geomech*, Balkema, 3-13.

Baise, L.G., Dreger, D.S., Glaser S.D. 2003. The effect of shallow San Francisco bay sediments on waveforms recorded during the Mw 4.6 Bolinas California Earthquake, *Bulletin of the Seismological Society of America*, 93, 465-479.

Bard, P.Y., Riepl-Thomas, J. 2000. Wave propagation in complex geological structures and their effects on strong ground motion, *Wave motion in earthquake eng.*, Kausel & Manolis eds, WIT Press, Southampton, Boston, 37-95.

Bard, P.Y., Bouchon, M. 1985. The two dimensional resonance of sediment filled valleys, *Bulletin of the Seismological Society of America*, 75, 519-541.

Basu U., Chopra A.K. 2003. Perfectly matched layers for time-harmonic elastodynamics of unbounded domains: theory and finite-element implementation, *Computer Meth. in Applied Mech. and Eng.*,192, 1337-1375.

Beauval, C., Bard, P.Y., Moczo, P., Kristek, J. 2003. Quantification of frequency-dependent lengthening of seismic ground-motion duration due to local geology: applications to the Volvi area (Greece), *Bull. Seism. Soc. of America*, 93, 371-385.

Beskos, D.E. 1997. Boundary elements methods in dynamic analysis: Part II (1986-1996), *Appl. Mech. Rev. (ASME)*, 50(3), 149-197.

Bielak, J., Xu, J., Ghattas, O. 1999. Earthquake ground motion and structural response in alluvial valleys, *Jal of Geotechnical and Geoenvironmental Eng.*, 125, 413-423.

Bielak J., Loukakis K., Hisada Y., Yoshimura C. 2003. Domain Reduction Method for Three-Dimensional Earthquake Modeling in Localized Regions, Part I: Theory, *Bulletin of the Seismological Society of America*, 93 (2), 817-824.

Bohlen, T. and E. H. Saenger, 2006. Accuracy of heterogeneous staggered-grid finite-difference modeling of rayleigh waves: *Geophysics*, 71, 109–115.

Bonilla F. 2000. *Computation of Linear and Nonlinear Site Response for Near Field Ground Motion*, Ph.D., University of California at Santa Barbara.

Bonnefoy-Claudet, S., Cornou, C., Bard, P.-Y., Cotton, F., Moczo, P., Kristek, J., Faeh, D., 2006. H/V ratio: a tool for site effects evaluation. Results from 1D noise simulations., *Geophys. J. Int.*, 167, 827-837.

Bonnet, M. 1999. *Boundary integral equation methods for solids and fluids*. Wiley.

Bonnet, M., Maier, G., Polizzotto, C. 1998. Symmetric Galerkin boundary element methods. *ASME Appl. Mech. Rev.*, 51(11), 669-704.

Bonnet, M., Bui H.D. 1993. Regularization of the displacement and traction BIE for 3D elastodynamics using indirect methods. *Advances in Boundary Element Techniques*, Kane JH, Maier G., Tosaka N. and Atluri SN (eds), Springer, Berlin, 1-29.

Bouchon, M., Campillo, M. and Gaffet, S. 1989. A Boundary Integral Equation-Discrete Wavenumber Representation Method to Study Wave Propagation in Multilayered Media Having Irregular Interfaces, *Geophysics*, 54, 1134-1140.

Bouchon, M. 1973. Effects of topography on surface motion, *Bull. Seismological Society of America*, 63, 615-622.

Bouckovalas G.D., Papadimitriou A.G. 2005. Numerical evaluation of slope topography effects on seismic ground motion, *Soil Dynamics and Earthquake Eng.*, 25(7-10), 547-558.

Brennan A.J., Madabhushi S.P.G. 2002. Effectiveness of vertical drains in mitigation of liquefaction, *Soil Dynamics and Earthquake Eng.*, 22(9-12), 1059-1065.

Carcione J.M., Cavallini F., Mainardi F., Hanyga A. 2002. Time-domain Modeling of Constant-Q Seismic Waves Using Fractional Derivatives, *Pure and Applied Geoph.*, 159(7-8), 1719-1736.

Chadwick E, Bettess P and Laghrouche O. 1999. Diffraction of short waves modelled using new mapped wave envelope finite and infinite elements. *Int. Journal for Numerical Methods in Eng.*, 45, 335-354.

Chaillat S., Bonnet M., Semblat J.F. 2009. A new fast multi-domain BEM to model seismic wave propagation and amplification in 3D geological structures, *Geophysical Journal International*, 177, 509-531.

Chaillat S., Bonnet M., Semblat J.F. 2008. A multi-level Fast Multipole BEM for 3-D elastodynamics in the frequency domain, *Computer Methods in Applied Mechanics and Engineering*, 197(49-50), 4233-4249.

Chaljub E., Komatitsch D., Vilotte J.P., Capdeville Y., Valette B., Festa G. 2007. Spectral Element Analysis in Seismology, in *Advances in Wave Propagation in Heterogeneous Media*, edited by Ru-Shan Wu and Valérie Maupin, Advances in Geophysics, Elsevier, 48, 365-419.

Chammas R., Abraham O., Cote P., Pedersen H., Semblat J.F. 2003. Characterization of heterogeneous soils using surface waves : homogenization and numerical modeling, *International Journal of Geomechanics* (ASCE), 3(1), 55-63.

Chávez-Garcia, F.J., Cardenas-Soto, M. 2002. The contribution of the built environment to the 'free-field' ground motion in Mexico city, *Soil Dynamics and Earthquake Eng.*, 22, 773-780.

Chávez-García, F.J., Raptakis, D.G., Makra, K., Pitilakis, K.D. 2000. Site effects at Euroseistest-II. Results from 2D numerical modelling and comparison with observations, *Soil Dynamics and Earthquake Eng.*, 19(1), 23-39.

Chávez-García, F.J., Stephenson, W.R., Rodríguez, M. 1999. Lateral propagation effects observed at Parkway, New Zealand. A case history to compare 1D vs 2D effects, *Bull. Seismological Society of America*, 89, 718-732.

Chopra A.K. 2007. *Dynamics of Structures, Theory and Applications to Earthquake Eng.*, Pearson Prentice Hall, 3rd Ed., 876p.

Clouteau, D., Aubry, D. 2001. Modifications of the ground motion in dense urban areas, *Journal of Computational Acoustics*, 9, 1659-1675.

Dangla P., Semblat J.F., Xiao H., Delépine N.., A simple and efficient regularization method for 3D BEM: application to frequency-domain elastodynamics, *Bull. Seism. Soc. Am.*, 95, 2005, 1916-1927.

Dangla, P. 1988. A plane strain soil-structure interaction model. *Earthquake Engineering and structural Dynamics*, 16, 1115-1128.

Darve E. 2000. The fast multipole method: Numerical implementation, *J. Comp. Phys.*, 160, 195-240.

Dauksher W., Emery A.F. 1999. An evaluation of the cost effectiveness of Chebyshev spectral and p-finite element solutions to the scalar wave equation, *Int. Jal for Numerical Methods in Eng.*, 45, 1099-1113.

Day S.M., Minster J.B. 1984. Numerical simulation of wavefields using a Padé approximant method, *Geophys. J. Roy. Astr. Soc.*, 78, 105-118.

Delépine N., Bonnet G., Semblat J.F., Lenti L. 2007, A simplified non linear model to analyze site effects in alluvial deposits, *4th Int. Conf. on Earthquake Geotechnical Eng.*, Thessaloniki, Greece.

Delépine N., Lenti L., Bonnet G., Semblat J.F. 2009. Nonlinear viscoelastic wave propagation: an extension of Nearly Constant Attenuation (NCQ) models, *Jal of Engineering Mechanics (ASCE)*, to appear.

Deraemaeker A., Babuška, I., Bouillard P. 1999. Dispersion and pollution of the FEM solution for the Helmholtz equation in one, two and three dimensions, *Int. Journal for Numerical Methods in Engineering*, 46, 471-499.

Dravinski, M. and Mossessian, T.K. 1987. Scattering of Plane Harmonic P, SV and Rayleigh Waves by Dipping Layers of Arbitrary Shape, *Bulletin of the Seismological Society of America*, 77, 212-235.

Emmerich H., Korn M. 1987. Incorporation of attenuation into time-domain computations of seismic wave fields. *Geophysics*, (59) 9, 1252-1264.

Faccioli, E., Maggio, F., Quarteroni, A. and Tagliani, A. 1996. Spectral Domain Decomposition Methods for the Solution of Acoustic and Elastic Wave Equations, *Geophysics*, 61, 1160-1174.

Festa G., Nielsen S. 2003. PML absorbing boundaries, *Bulletin of the Seismological Society of America*, 93(2), 891-903.

Festa, G. and Vilotte J.-P. 2005. The Newmark scheme as a Velocity-Stress Time staggering: An efficient PML for Spectral Element simulations of elastodynamics, *Geophys. J. Int.*, 161, 3, 789-812.

Frankel, A. and Vidale, J. 1992. A Three-Dimensional Simulation of Seismic Waves in the Santa Clara Valley, California, from a Loma Prieta Aftershock, *Bulletin of the Seismological Society of America*, 82, 2045-2074.

Fujiwara H. 2000. The fast multipole method for solving integral equations of three-dimensional topography and basin problems, *Geophysical Journal International*, 140, 198-210.

Gaul L., Schanz M. 1999. A comparative study of three boundary element approaches to calculate the transient response of viscoelastic solids with unbounded domains, *Computer Methods in Applied Mechanics and Eng.*, 179(1-2), 111-123.

Glinsky-Olivier N., Benjemaa M., Piperno S., Virieux J. 2006. A finite-volume method for the 2D seismic wave propagation, *Europ. Geophysical Union General Assembly*, Vienna, Austria, April 2-7.

Graves R.W. 1996. Simulating seismic wave propagation in 3D elastic media using staggered-grid finite differences, *Bulletin of the Seismological Society of America*, 86, 1091-1106.

Greengard L., Huang J., Rokhlin V., and Wandzura S. 1998. Accelerating fast multipole methods for the Helmholtz equation at low frequencies. *IEEE Comp. Sci. Eng.*, 5(3), 32-38.

Groby J.P., Tsogka C., Wirgin A. 2005. Simulation of seismic response in a city-like environment, *Soil Dynamics and Earthquake Eng.*, 25(7-10), 487-504.

Guiggiani, M. 1994. Hypersingular formulation for boundary stress evaluation. *Eng. Analysis Boundary Elements*, 13, 169-179.

Gyebi O.K., Dasgupta G. 1992. Finite element analysis of viscoplastic soils with Q-factor, *Soil Dynamics and Earthquake Eng.*, 11(4), 187-192.

Heuze, F., Archuleta R., Bonilla F., Day S., Doroudian M., Elgamal A., Gonzales S., Hoehler M., Lai T., Lavallee D., Lawrence B., Liu P.C. 2004. Estimating site-specific strong earthquake motion, *Soil Dynamics and Earthquake Eng.*, 24, 199-223.

Hughes, T.J.R. 1987. *Linear static and dynamic finite element analysis*, Prentice-Hall, Englewood Cliffs, N-J.

Hughes T.J.R., Reali A., Sangalli G. 2008. Duality and Unified Analysis of Discrete Approximations in Structural Dynamics and Wave Propagation: Comparison of p-method Finite Elements with k-method NURBS, *Computer Meth. in Applied Mech. and Eng.*, 197(49-50), 4104-4124.

Ihlenburg, F., Babuška, I. 1995. Dispersion analysis and error estimation of Galerkin finite element methods for the Helmholtz equation, *Int. Journal for Numerical Methods in Engineering*, 38, 3745-3774.

Irikura K., Kamae K. 1994. Estimation of strong ground motion in broad-frequency band based on a seismic source scaling model and and empirical Green's function technique, *Annali di Geofisica*, XXXVII-6, 1721-1743.

Jin, F., Pekau, O.A., Zhang, C.H. 2001. A 2D time-domain boundary element method with damping, *Int. Journal for Numerical Methods in Engineering*, 51, 647-661.

Kausel E., Assimaki, D. 2002. Seismic simulation of inelastic soils via frequency-dependent moduli and damping, *Jal of Eng. Mechanics*, 128(1), 34-47.

Kham M., Semblat J.F., Bard P.Y., Dangla P. 2006. Seismic Site-City Interaction: Main Governing Phenomena through Simplified Numerical Models, *Bulletin of the Seismological Society of America*, 96(5), 1934-1951.

Kjartansson E. 1979. Constant Q wave propagation and attenuation, *J. Geophys. Res.*, 84, 4737-4748.

Komatitsch, D. and Vilotte, J.P. 1998. The Spectral Element Method: An Efficient Tool to Simulate the Seismic Response of 2D and 3D Geological Structures, *Bulletin of the Seismological Society of America*, 88, 368-392.

Lade P.V. 1977. Elasto-plastic stress-strain theory for cohesionless soil with curved yield surfaces, *Int. Journal of Solids and Structures*, 13, 1019-1035.

Liao, W.I., Teng, T.J., Yeh, C.S. 2004. A series solution and numerical technique for wave diffraction by a three-dimensional canyon, *Wave Motion*, 39, 129-142.

Lombaert, G., Clouteau, D., Ishizawa, O., Mezher, N. 2004. The city-site effect : a fuzzy substructure approach and numerical simulations, *11th Int. Conf. on Soil Dynamics and Earthquake Eng.*, University of California at Berkeley, January 7-9.

Loret B., Simões F.M.F., Martins J.A.C. 1997. Growth and decay of acceleration waves in non-associative elastic-plastic fluid-staturated porous media, *Int. Jal of Solids and Structures*, 34(13), 1583-1608.

Manolis G.D., Beskos D.E. 1988. *Boundary Element Methods in Elastodynamics*. Unwin-Hyman (Chapman & Hall), London.

Mellal A., Modaressi H. 1998. A simplified numerical approach for nonlinear dynamic analysis of multilayered media, *11th European Conf. on Earthquake Eng.*, Balkema ed., Paris, France.

Meza-Fajardo K.C., Papageorgiou A.S. 2008. A Nonconvolutional, Split-Field, Perfectly Matched Layer for Wave Propagation in Isotropic and Anisotropic Elastic Media: Stability Analysis, *Bulletin of the Seismological Society of America*, 98(4), 1811-1836.

Mikhailenko, B.G. 1999. Spectral Laguerre method for the approximate solution of time dependent problems, *Appl. Math. Letters*, 12, 105-110.

Moczo P., Kristek J. 2005. On the rheological models used for time-domain methods of seismic wave propagation, *Geophys. Res. Lett.*, 32, L01306.

Moczo, P., Kristek, J., Vavrycuk, V., Archuleta, R.J., Halada, L. 2002. 3D Heterogeneous Staggered-Grid Finite-Difference Modeling of Seismic Motion with Volume Harmonic and Arithmetic Averaging of Elastic Moduli and Densities, *Bulletin of the Seismological Society of America*, 92(8), 3042-3066.

Modaressi, H., Benzenati, I. 1994. Paraxial Approximation for Poroelastic Media, *Soil Dynamics and Earthquake Engineering*, 13, pp. 117-129.

Paolucci, R. 2002. Amplification of earthquake ground motion by steep topographic irregularities, *Earthquake Eng. and Structural Dynamics*, 31, 1831-1853.

Paolucci, R. 1999. Shear resonance frequencies of alluvial valleys by Rayleigh's method, *Earthquake Spectra*, 15, 503-521.

Park D., Hashash Y.M.A. 2004. Soil damping formulation in nonlinear time domain site response analysis, *Jal of Earthquake Eng.*, 8(6), 249-274.

Pavlenko, O. 2001. Nonlinear seismic effects in soils: numerical simulation and study, *Bull. Seism. Soc. Am.*, 91(2), 381-396.

Pitilakis, K.D., Raptakis, D.G., Makra, K.A. 1999. Site effects : recent considerations and design provisions, *2nd Int. Conf. on Earthquake Geotechnical Eng.*, Lisbon, Balkema ed, 901-912.

Prevost J.H. 1985. A simple plasticity theory for frictional cohesionless soils, *Soil Dynamics and Earthquake Eng.*, 4(1), 9-17.

Reinoso E., Wrobel L. C., Power H. 1997.Three-dimensional scattering of seismic waves from topographical structures, *Soil. Dyn. Earthquake Engng.*, 16, 41–61.

Sánchez-Sesma, F.J., Vai, R., Dretta, E., Palencia, V.J. 2000. Fundamentals of elastic wave propagation for site amplification studies, *Wave Motion in Earthquake Engineering*, E. Kausel and G. Manolis (Editors), WIT Press, Southampton, UK, 1-36.

Sanchez-Sesma, F.J. and Luzon, F. 1995. Seismic Response of Three-Dimensional Alluvial Valleys for Incident P, S and Rayleigh Waves, *Bulletin of the Seismological Society of America*, 85, 269-284.

Sanchez-Sesma, F.J. 1983. Diffraction of elastic waves by three-dimensional surface irregularities, *Bulletin of the Seismological Society of America*, 73(6), 1621-1636.

Schnabel P.B., Lysmer J., Seed H.B. 1972. *SHAKE-A computer program for equation response analysis of horizontally layered sites*, Rep. No. EERC 72-12, University of California, Berkeley.

Seed H.B., Wong R.T., Idriss I.M., Tokimatsu K. 1986. Moduli and damping factors for dynamic analyses of cohesionless soils, *Jal of Geotechnical Eng.*, 112(11), 1016-1032.

Semblat J.F., Pecker A. 2009. *Waves and Vibrations in Soils: Earthquakes, Traffic, Shocks, Construction Works*, IUSS Press, Pavia, Italy, 500 p.

Semblat J.F., Kham M., Bard P.Y. 2008. Seismic wave propagation in alluvial basins and influence of site-city interaction, *Bulletin of the Seismological Society of America*, 98(6), 2665-2678.

Semblat J.F., Kham M., Parara E., Bard P.Y., Makra K., Raptakis D. 2005. Site effects: basin geometry vs soil layering, *Soil. Dyn. Earthquake Eng.*, 25, 529-538.

Semblat J.F., Paolucci R., Duval AM. 2003. Simplified vibratory characterization of alluvial basins, *C. R. Geoscience*, 335, 365-370.

Semblat, J.F., Duval, A.M., Dangla P. 2002. Seismic Site Effects in a Deep Alluvial Basin: Numerical Analysis by the boundary element method, *Computers and Geotechnics*, 29(7), 573-585.

Semblat J.F., Brioist J.J. 2000a. Efficiency of higher order finite elements for the analysis of seismic wave propagation, *Journal of Sound and Vibration*, 231(2), 460-467.

Semblat, J.F., Duval, A.M., Dangla P. 2000b. Numerical analysis of seismic wave amplification in Nice (France) and comparisons with experiments, *Soil Dynamics and Earthquake Eng.*, 19(5), 347-362.

Semblat J.F. 1997. Rheological interpretation of Rayleigh damping, *Journal of Sound and Vibration*, 206(5), 741-744.

Sextos A.G., Pitilakis K.D., Kappos A.J. 2003. Inelastic dynamic analysis of RC bridges accounting for spatial variability of ground motion, site effects and soil-structure interaction phenomena. Part 1: Methodology and analytical tools, *Earthquake Engineering and Structural Dynamics*, 32(4), 607-627.

Sladek V., Sladek J. 1998. Singular integrals and boundary elements. *Comput. Methods Appl. Mech. Engrg.*, 157, 251-266.

Virieux J. 1986. P-SV Wave Propagation in Heterogeneous Media: Velocity-Stress Finite-Difference Method, *Geophysics*, 51, 889-901.

Yokoi T. 2003. The higher order Born approximation applied to improve the solution of seismic response of a three-dimensional canyon by the Indirect Boundary Method. *Physics of the Earth and Planetary Interior*, 137, 97-106.

Sommerville P.G. 1998. Emerging art: earthquake ground motion, ASCE Geotechnical Special Publications, Dakoulas et al eds, 1, 1-38.

Takahashi A., Takemura J. 2005. Liquefaction-induced large displacement of pile-supported wharf, *Soil Dynamics and Earthquake Eng.*, 25, 11, pp.811-825.

Takahashi T., Nishimura N., Kobayashi S. 2003. A fast BIEM for three-dimensional elastodynamics in time domain, *Engineering Analysis with Boundary Elements*, 27, 491-506.

Tsogka C., Wirgin A. 2003. Seismic response of a set of blocks partially embedded in soft soil, *C.R.Mecanique*, 331(3), 217-224.

Wirgin A., Bard, P-Y., 1996. Effects of buildings on the duration and amplitude of ground motion in Mexico city, *Bulletin of the Seismological Society of America*, 86, 914-920.

Wolf J.P., 2003. *The scaled boundary finite element method*, Wiley, Chichester, UK.